\documentclass{PoS}

\usepackage{journals}
\usepackage{natbib}
\title{The Boundary Layer in compact binaries}

\ShortTitle{The BL in compact binaries}

\author{\speaker{Marius Hertfelder}\\
        Institut f\"ur Astronomie und Astrophysik, Abt. Computational Physics\\
        Auf der Morgenstelle 10, 72076 T\"ubingen, Germany\\
        E-mail: \email{marius.hertfelder@gmail.com}}

\abstract{
The structure and the
thermodynamics of the non-magnetic boundary layer (BL) of accretion disks
has been an outstanding problem in the field of theoretical astrophysics for
years. The BL is a ubiquitous phenomenon that appears in a variety of
astrophysical situations and systems where non-magnetic accretion occurs,
i.e. where an accretion disk (AD) is present. The AD is an efficient mechanism to
transport matter from the exterior of the disk to the gravitating center.
Here, at the inner edge of the AD, the circulating matter comes upon the
surface of the central object and is decelerated to match the object's
rotation rate. During this process, an enormous amount of energy is released
from the tiny BL region. This in turn generates hard radiation which can be clearly
identified in the observed spectrum of the object. We perform numerical
hydrodynamical simulations in order to calculate the luminosity and the
spectrum of the BL and its dependence on parameters like the mass, rotation
rate or mass accretion rate of the central white dwarf (WD). Therefore, we
treat the problem in the one-dimensional, radial slim disk approximation. We
employ a classical $\alpha$-viscosity to account for the turbulence and
include cooling from the disk surfaces as well as radial radiation transport.
To account for the high temperatures in BLs around WDs, we also consider the
radiation energy in a one-temperature approximation. We find that 1D models
of the BL are well suited if one is interested in the radiation characteristics
of the BL. The BL luminosity directly depends on the varied parameters which
makes it possible to draw conclusions about real systems by comparing observations
with our synthetic models. Ambiguities concerning different models with identical
luminosities can be mitigated by regarding the emitted spectrum. We therefore
present a method to gain information about a system by probing the radiation
of the BL.
}

\FullConference{The Golden Age of Cataclysmic Variables and Related Objects - III, Golden2015 \\
		7-12 September 2015\\
		Palermo, Italy}

\begin{document}

\section{Introduction}

Cataclysmic variables (CVs) are binary systems which are of great interest within the context of accretion physics since one of their main features
is the mutual exchange of mass \citep{1971ARA&A...9..183P}. This leads to the formation of an accretion disk around a roughly solar mass white dwarf
(WD) which is fueled by the Roche lobe overflow of the lighter main sequence companion star \citep{1995CAS....28.....W,giovannelli1985multifrequency}.
The accretion disk is an efficient mechanism for the matter to get rid of its angular momentum and travel towards the central WD. During this journey,
energy is released due to the fact that the matter is falling deeper into the gravitational potential of the WD \citep[e.g.][]{1973A&A....24..337S,1974MNRAS.168..603L,1981ARA&A..19..137P,1982SSRv...32..379V}.
Over the whole extent of the disk, about one half of the accretion energy, which is given by
\begin{equation}
	L_\mathrm{acc} = \frac{GM_\ast\dot{M}}{R_\ast}\label{eq:Lacc}
\end{equation}
($G, M_\ast, \dot{M}$ and $R_\ast$ are the gravitational constant, WD mass, mass accretion rate and radius) for non-rotating stars, becomes available.
The other half of the energy specified by Eq.~(\ref{eq:Lacc}) is stored in terms of kinetic energy of the gas which rotates with Keplerian velocity
$\Omega_\mathrm{K}=\frac{GM_\ast}{r^3}$ near the surface of the WD. In order to match the rotation rate of the WD, which is in general much slower or
even zero, the gas must be strongly slowed down before it can settle on the stellar surface. During this deceleration, an enormous amount of energy is
released in a spatially confined region which is called the BL and has a radial extent of around one percent of the stellar radius for the
case of a WD. The resulting UV and soft and hard X-ray emission of these hot BLs has been observed in several CVs \citep[e.g.][]{1981MNRAS.196....1C,1981ApJ...245..609C,1984MNRAS.206..879C}.
Depending on the mass accretion rate of the WD, the BL can either be optically thin ($\dot{M}\leq 10^{-10} M_{\odot}/\mathrm{yr}$, \citealt{1987MNRAS.227...23W})
and the radiation will be dominated by soft and hard X-rays (e.g. \citealt{2004RMxAC..20..244M,2003MNRAS.346.1231P,2005ApJ...626..396P},
\citealt{1984Natur.308..519K,1987Ap&SS.130..303S,1993Natur.362..820N,1999MNRAS.308..979P}), or it can be optically thick and emit thermal radiation
(see e.g. \citealt{1980MNRAS.190...87C,2004RMxAC..20..174M}).

For nearly 50 years now, it has been the goal of many astrophysicists to theoretically reproduce the BL. There have been several approaches to accomplish
this task, one of the first being stationary calculations or timescale estimates \citep{1974MNRAS.168..603L,1977MNRAS.178..195P,1977AcA....27..235T,1981AcA....31..267T,1979MNRAS.187..777P,1983A&A...126..146R}.
With increasing computational power, the era of numerical hydrodynamics was introduced and the first evolutionary calculations have been performed
\citep{1986MNRAS.221..279R,1987A&A...172..124K,1989A&A...208...98K,1989A&A...222..141K,1991A&A...247...95K,1995MNRAS.275.1093G}.
The latter authors used a one-dimensional approximation of the BL, which is still a viable approach for certain aims. The gas is assumed to be slowed
down in the midplane of the disk before it is spread on the surface of the star. Within this model, considering only the radial dependence of the physical
variables, is then sufficient for the calculation of the total radiation emerging from the BL. A modern version of this approach has been presented in
\citet{2013A&A...560A..56H} where we included a quasi-two-dimensional radiation transport and special treatment for the radiation field.

Some questions about the BL cannot be answered by the 1D approximation of the BL and therefore, multidimensional models have been pursued as well.
Among the first full radiation hydrodynamical simulations were the efforts by \citet{1989A&A...208...98K,1989A&A...222..141K,1991A&A...247...95K},
who did two-dimensional $r$-$\vartheta$-simulations assuming axisymmetry. Those simulations are apt to investigate the structure of the BL and the fate of the disk material, i.e.
the meridional spreading and the mixing with the stellar material. With the availability of large compute clusters and parallel hydrodynamics codes,
simulations of this kind nowadays feature amazing numerical resolutions and long evolution times, while including sophisticated physics and an advanced
treatment of radiation (see Hertfelder \& Kley 2017, in preparation). However, two-dimensional simulations in the disk plane ($r$-$\varphi$)
were of great interest in the recent years. In order to slow down the gas in the BL,
some mechanism of angular momentum (AM) transport must be present. In wide parts of the disk, this is done by the magnetorotational
instability (MRI), which creates turbulence that acts like a genuine viscosity on macroscopic scales. In this case, the viscosity can be considered
using the classical $\alpha$-prescription by \citet{1973A&A....24..337S}. In the BL, though, the source of the observed AM transport is still a
matter of ongoing research. Recent simulations revealed the existence of a supersonic instability in the BL, which excites acoustic waves that are able to
transport AM and mass \citep[see][for details]{2015A&A...579A..54H}. Full three-dimensional simulations have been done sporadically \citep[e.g.][]{2002MNRAS.330..895A},
however, the computational costs for highly resolved models are immense and the simulation times are very long.

In this work, we focus on the BL around a solar mass WD in a cataclysmic variable system. We extent the study presented in \citet{2013A&A...560A..56H}
and analyze the luminosity and the compound black body spectrum of the BL as a function of important system parameters such as the mass accretion
rate and the stellar rotation rate. Detailed spectra, that also take the vertical structure into account, have been presented in \citet{2014a&a...571a..55s}.

\section{Model \& Physics}\label{sec:model}

The problem is approached in a one-dimensional approximation in a cylindrical coordinate system ($r$, $\varphi$, $z$). For this purpose, the Navier-Stokes
equations have been integrated in the vertical direction and derivatives with respect to the azimuthal direction $\varphi$ have been dropped due
to the assumption of axisymmetry. This approximation is called the \emph{thin disk} approach and the variables depend only on radius and time ($r$, $t$).
The mass density $\rho$ is replaced by a vertically integrated surface density $\Sigma$ which can be derived by
\begin{equation}
	\Sigma = \int_{-\infty}^{\infty} \rho\mathrm{d}z = \sqrt{2\pi}\rho(z=0)H,\label{eq:sigma}
\end{equation}
where we assumed a Gaussian profile for $\rho$ in the vertical direction. $H$ is the pressure scale height and thus a measure for the height of the disk.
Assuming hydrostatic balance and an isothermal equation of state in $z$-direction, it reads
\begin{equation}
	H = \frac{c_\mathrm{s}}{\Omega_\mathrm{K}},
\end{equation}
where $c_\mathrm{s}$ is the sound speed. In the radial direction we use the ideal gas law for the pressure.

Since radiation pressure and energy are not negligible due to the high temperatures in the BL, we employ the one-temperature radiation transport
\citep[see e.g.][]{2010MNRAS.409.1297F}, where the two equations for the gas and the radiation energy density are added up. We then propagate
the total energy, consisting of gas and radiation energy, in time. This approach is a good approximation for optically thick regions \citep[e.g.][]{2010A&A...511A..81K}
and justified since the BLs we are regarding here are optically thick. The radiation energy equation is closed by employing the flux-limited
diffusion approximation (FLD; \citet{1981ApJ...248..321L,1984JQSRT..31..149L}) for the radiative flux $\vec{F}$ and we adopt the formulation by
\citet{1981ApJ...248..321L} for the flux-limiter $\lambda$. The opacity $\kappa$ is determined using Kramer's law,
\begin{equation}
	\kappa = 5\times10^{24} \rho T^{-3.5} \,[\mathrm{cm}^2\mathrm{g}^{-1}],
\end{equation}
with a lower threshold given by Thomson scattering. The disk can cool vertically via a blackbody radiation of temperature $T_\mathrm{eff}$, which
is calculated from the midplane temperature $T$ by using a generalization of the gray atmosphere for the optical depth in the vertical direction
\citep{1990ApJ...351..632H}. The exact equations for our model can be found in \citet{2013A&A...560A..56H}.

The partial differential equations are discretized on a fixed Eulerian grid using finite differences. For the time propagation, a semi-implicit-explicit
scheme is employed since some source terms, especially the viscous ones, require an implicit treatment so that the time step is not restricted too
severely. The code \citep[see also][]{2013A&A...560A..56H} maintains a formal second-order accuracy in time and space and uses a multi step procedure
for the time integration (operator splitting), which is controlled by the CFL condition that limits the largest possible time step.

The boundary conditions are implemented such that the disk is fed from the outer radius with a constant rate $\dot{M}$. We impose Keplerian rotation
at the outer edge and stellar rotation $\Omega_\ast$ at the inner edge. For the other variables, we assume zero gradient boundary conditions.
The models are started from initial profiles given by the disk solution by \citet{1973A&A....24..337S} which have been interpolated to the stellar
surface.

\subsection{Model Parameters}

We focus on the BL around a WD in a cataclysmic variable system. Accordingly, simulations for a WD with $0.6, 0.8$ and $1.0$
solar masses have been performed. We assumed different values for the mass accretion rate $\dot{M}$ that range from $10^{-10}$ up to $10^{-8}$
solar masses per year. As can be seen from Eq.~(\ref{eq:Lacc}), another crucial parameter for the luminosity of the BL is the stellar radius
$R_\ast$. For WDs, mass and radius are not independent but connected via an inverse relation; i.e. the larger the stellar mass, the smaller the
radius. We use the relation from \citet{1972ApJ...175..417N} to determine the exact value of $R_\ast$. The stellar rotation rate also plays
an important role for the luminosity of the BL. We varied it between $0.0$ and $0.9 \Omega_\mathrm{K}(R_\ast)$, which corresponds to non-rotating
up to nearly break-up velocity. The parameter for the $\alpha$-viscosity was taken to be $0.01$ throughout.

\section{Results}

\subsection{The general structure of the BL}

We will begin our discussion of the results by first presenting the basic properties of the BL. For this purpose we adduce the case with $M_\ast=0.8M_\odot$
and $\dot{M}=10^{-8} M_\odot/\mathrm{yr}$ as a standard model and basis for the parameter variations.

\begin{figure}[t]
\begin{center}
\includegraphics[width=\textwidth]{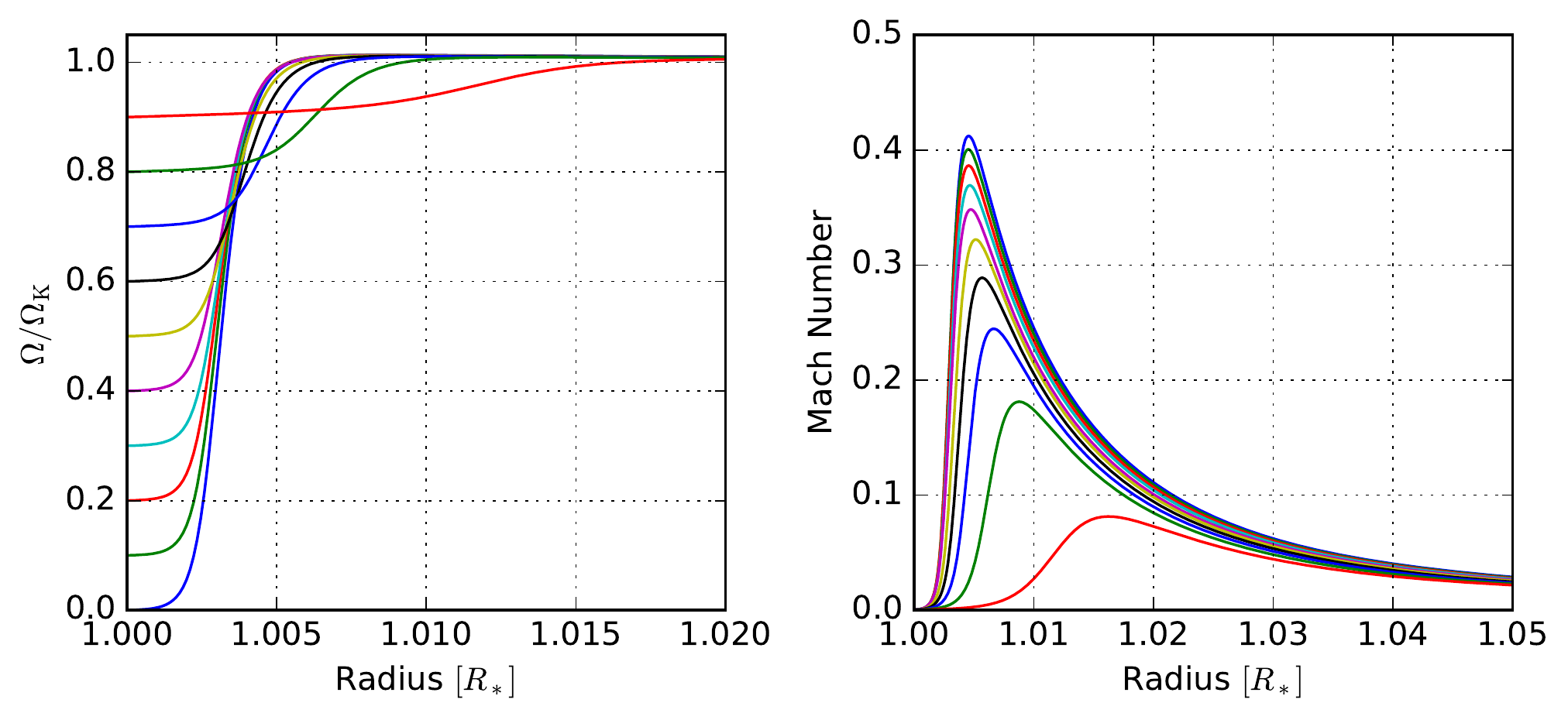}
\caption{\label{fig:Om-Ma}%
Angular velocity $\Omega=v_\varphi/r$ in terms of the Keplerian angular velocity (\emph{left panel}) and radial Mach number $\mathrm{Ma} = - v_r/c_\mathrm{s}$ 
(\emph{right panel}) as a function of the radial coordinate in units of $R_\ast$. The different colors denote the different stellar rotation rates
$\Omega_\ast$ which can be inferred from the left panel at $r=1$.
}
\end{center}
\end{figure}

Figure~\ref{fig:Om-Ma} shows the dynamical structure of the BL. On the left hand side, the angular velocity divided by the Keplerian angular velocity
is depicted in the region between the stellar surface and the disk. Viewing the situation from the outside to the inside, the gas initially rotates with Keplerian
velocity in the disk due to the force balance of gravitational and centrifugal forces. There is a small, additional pressure support since the temperature
decreases with increasing radius and thus the gas may rotate sub-Keplerian. This situation changes when the pressure gradient is pointing
inwards as we come closer to the BL. Then, the gas rotates slightly super-Keplerian. When entering the BL, the gas reaches the maximum rotational
velocity, a point which is called the zero-torque point, since no viscous torques exist at this point due to the vanishing gradient of $\Omega$. Going further
inwards, the gas is decelerated smoothly down to the velocity of the stellar surface. Here, it is mainly stabilized by pressure support. The region between the stellar surface and the maximum of $\Omega$ is
denoted by the \emph{dynamical BL width} and there is a tendency for an increasing BL width with increasing stellar rotation rate.

The radial or infall velocity of the matter is reflected by the Mach number shown in the right panel of Fig.~\ref{fig:Om-Ma}. Approaching the BL from the outside,
the infall velocity of the gas increases and reaches its maximum at the zero-torque point before it drops rapidly to almost zero at the surface of the star.
Thus, there is no shocking of the gas at the stellar surface. The increasing radial velocity is due to the loss of angular momentum caused by friction in
the disk and the slower the star spins, the higher is the infall velocity. Apparently no supersonic infall velocities are reached, which is an important issue in connection with the
causality \citep[see e.g.][]{1977MNRAS.178..195P,1992ApJ...394..255P,1997MNRAS.285..239K}. In general, we do not find supersonic infall velocities in any of our models
for small values of $\alpha$.

\begin{figure}[t]
\begin{center}
\includegraphics[width=\textwidth]{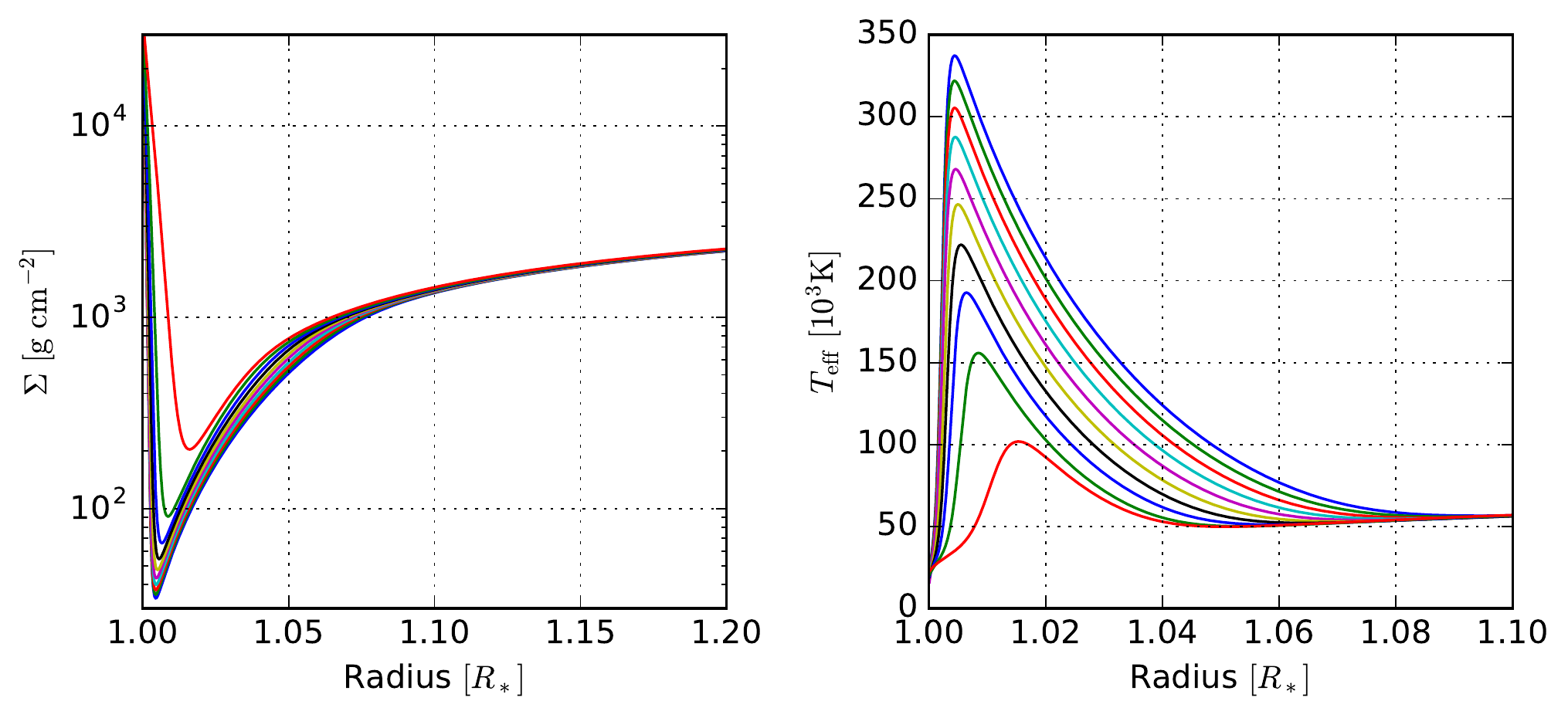}
\caption{\label{fig:Sigma-Teff}%
Surface density $\Sigma$ (\emph{left panel}) and effective or surface temperature $T_\mathrm{eff}$ (\emph{right panel}). The different
colors represent models with different stellar rotation rate and match those of Fig.~1.
}
\end{center}
\end{figure}

The thermal structure of the BL is illustrated in Fig.~\ref{fig:Sigma-Teff} with the surface density $\Sigma$ (see Eq.~\ref{eq:sigma}) on the left hand side
and the effective temperature on the right hand side. The color coding is analogue to Fig.~\ref{fig:Om-Ma}). The plot of the surface density shows that the
BL is heavily depleted of gas, $\Sigma$ decreases by almost two orders of magnitude compared to the disk. The BL can thus be seen as a bottleneck, where the
matter has to squeeze through in order to reach the surface of the star. This notion is in accordance with the behavior of the radial velocity: The smaller
the surface density becomes, the greater the radial velocity is in order to maintain a constant mass accretion rate,
\begin{equation}
	\dot{M} = -2\pi r\Sigma v_r.\label{eq:Mdot}
\end{equation}
Equation~\ref{eq:Mdot} can directly be derived from the conservation of mass. Furthermore, the depletion of mass in the BL depends on the stellar rotation
rate and is most severe for a non-rotating star. This does not apply for the disk where all models have about the same density. The rapid increase of $\Sigma$
at $r\approx1$ marks the beginning of the WD.

Although the physical model presented here is one-dimensional in nature, we can calculate a surface temperature from the midplane temperature by using
an appropriate approximation for the vertical structure (see Sec.~\ref{sec:model}). $T_\mathrm{eff}$ then represents the temperature of the disk and BL at
an optical depth of $\tau\approx1$ and thus determines the radiation emerging from the system. Figure~\ref{fig:Sigma-Teff} shows that the surface temperature
increases tremendously in the BL, reaching up to almost $350\,000$ Kelvin. This peak is due to the strong shearing in the BL where the gradient of $\Omega$
is large (see Fig.~\ref{fig:Om-Ma}) and a great deal of heat is produced through friction. With increasing stellar rotation rate, the maximum value of
$T_\mathrm{eff}$ in the BL shrinks, since the shearing becomes less intense. In the disk, there is no difference between the presented models. The
surface temperature shows a second peak at $r\approx1.41$ with $T_\mathrm{eff}=67152$ Kelvin which is in perfect agreement with the disk solution by
\citet{1973A&A....24..337S}. Although the region where the energy is produced has a radial extent of only about one percent of the stellar radius, the
major peak of $T_\mathrm{eff}$ spreads over almost 10 percent. The reason is the radial diffusion which transports energy
through the disk and the resulting region has been named the \emph{thermal BL} \citep{1995MNRAS.272...71R,1995ApJ...442..337P}. It is the area from which
the BL radiation that can be observed escapes.

%

\subsection{The width of the BL}

\begin{table}
	\begin{center}
		\begin{tabular}{c c c c}
		\hline\hline
		$\omega$ & $\Delta r$ $[R_\ast]$ \\\hline
		$0.0$ & $0.0071$ \\
		$0.1$ & $0.0071$ \\
		$0.2$ & $0.0071$ \\
		$0.3$ & $0.0071$ \\
		$0.4$ & $0.0072$ \\
		$0.5$ & $0.0076$ \\
		$0.6$ & $0.0081$ \\
		$0.7$ & $0.0089$ \\
		$0.8$ & $0.0112$ \\
		$0.9$ & $0.0182$ \\
		\hline
		\end{tabular}
	\end{center}
\caption{%
\label{tab:width}%
Width of the boundary layer for the standard model with $M_\ast=0.8 M_\odot$, $\dot{M}=10^{-8}M_\odot/\mathrm{yr}$ and stellar rotation rates $\omega=\Omega_\ast/\Omega_\mathrm{K}(R_\ast)$ spanning from $0.0$
up to $0.9$. By definition, the BL ranges from the surface of the star to the point where $\partial\Omega(r)/\partial r =0$, i.e. where it has a maximum.
}
\end{table}

Table~\ref{tab:width} shows the width of the dynamical BL for the reference model. It is defined to be the region ranging from the stellar surface, which
is at $r\approx1$ in our simulations, to the zero-gradient point where $\partial\Omega/\partial r=0$, i.e. the angular velocity has its maximum. This point
is not identical with the maximum point of the curves in the left panel of Fig.~\ref{fig:Om-Ma}. The width of the BL around a WD is extremely small with
values of less than one percent of the stellar radius in most cases. Only for fast rotating WDs is the BL becoming significantly wider. The reason for
the narrow BL is the large value of $M_\ast/R_\ast$ for WDs which have a mass comparable to our sun but a radius of only a hundredth of the solar radius. This
fraction appears in the gravitational force which is, among other parameters, responsible for the temperature and surface density of the disk and the BL.
The width is in general governed by the viscosity, which in turn depends on the surface density and the temperature if we employ a $\alpha$-prescription. Thus, the
high ratio of stellar mass and stellar radius in WDs leads to very thin BLs, as opposed to young stars, for instance. This is one of the reasons
why BL simulations of WDs are demanding from a computational point of view, since a high numerical resolution has to be applied in order to resolve
this small area sufficiently.

With increasing stellar rotation rate, the temperature in the BL decreases, since the gas retains more and more of its angular momentum and less kinetic energy
is released. Although a colder BL should be thinner than a hot one (one can think of the heat puffing up the BL), we observe the opposite, namely an increasing
BL width (see Table~\ref{tab:width}). The reason is that the depletion of mass in the BL depends on the rotation rate as well (see Fig.~\ref{fig:Sigma-Teff},
left panel), and $\Sigma$ is larger for high $\Omega_\ast$. The complex interplay of surface density and midplane temperature arranges it such that the BL
becomes wider. The thermal BL, however, is about ten times larger and diminishes in size with increasing stellar rotation due to the
decreasing energy release in the BL.

\begin{figure}[t]
\begin{center}
\includegraphics[width=\textwidth]{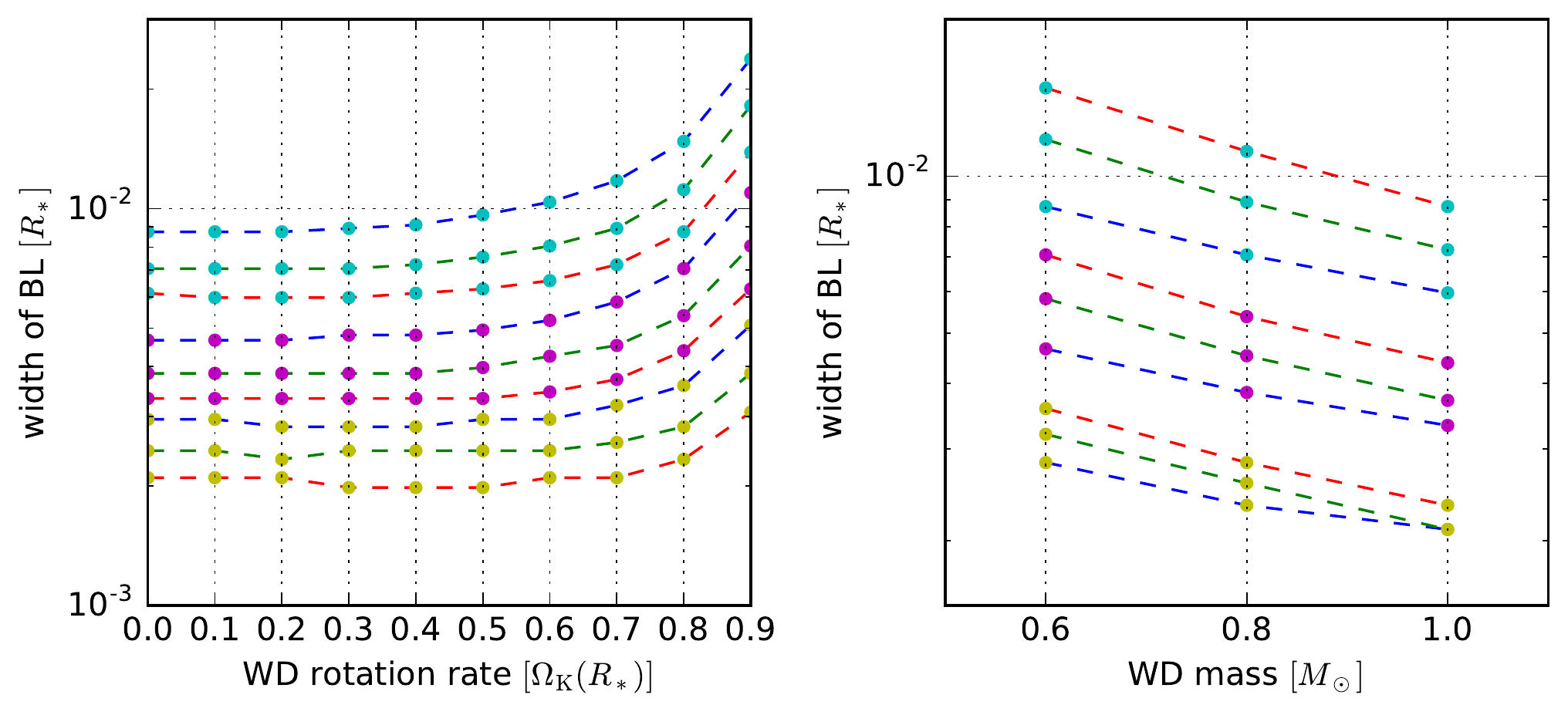}
\caption{\label{fig:width}%
The width of the dynamical BL as a function of the WD rotation rate (\emph{left panel}) and the WD mass (\emph{right panel}) in units of
the WD radius $R_\ast$. The marker color represents the mass accretion rate of the WD in both plots where cyan equals $10^{-8}$, magenta
$10^{-9}$ and yellow $10^{-10}$ solar masses per year. On the left hand side, the line color illustrates the mass of the WD with blue,
green and red being $0.6, 0.8$ and $1.0 M_\odot$. On the right hand side, the line color tells the rotation rate of the model and blue,
green and red correspond to $0.2$, $0.7$ and $0.9 \Omega_\mathrm{K}(R_\ast)$, respectively.
}
\end{center}
\end{figure}

Figure~\ref{fig:width} visualizes the width of the dynamical BL as a function of the WD rotation rate and mass for three different values
of the mass accretion rate $\dot{M}$. We will first refer to the graph on the left hand side, where the marker colors correspond to the
mass accretion rate and the line color represents the mass of the WD. The reference model is shown with cyan markers and a green dashed
line. As has been discussed in connection with Table~\ref{tab:width}, we recognize a clear trend for an increasing BL width with growing
stellar rotation rate. This is, however, not the case in some other models considered in Fig.~\ref{fig:width}. Especially for lower values
of the mass accretion rate, the BL might shrink in width before it is getting larger with increasing stellar rotation rate.
Consider, for instance, the bottom red line with yellow markers, which corresponds to $1.0 M_\odot, 10^{-10} M_\odot/\mathrm{yr}$, where
this behavior is visible. The reverse of the trend is due to the complex interplay of surface density and midplane temperature,
which has been mentioned before. Depending on the specific choice of parameters, the BL is either growing continually with
increasing rotation rate, or it is shrinking slightly in the beginning before it is widening again. It is impossible to give a general
answer to this issue, however, there seems to be a tendency for models with low mass accretion rate and high stellar mass to adopt the
latter behavior. Apart from this detail, all models show a similar overall trend: The width is increasing slowly in the beginning and
more severely for high stellar rotation rates. Thus, if one is able to identify the width of a BL (e.g. from the radiation characteristics),
it helps to distinguish between fast rotating WDs but not between low or non-rotating ones.

The right panel of Fig.~\ref{fig:width} shows how the mass of the WD influences the width of the BL. Again, the mass accretion rate is
visualized by the differently colored markers. We have picked three stellar rotation rates for each mass accretion rate and WD mass,
given by $0.2$ (blue line), $0.7$ (green line) and $0.8$ (red line) times the breakup velocity. Clearly, the BL is shrinking with increasing
WD mass. The reason is the increasing gravitational pull, which is enhanced even more by the inverse mass-radius relation for WDs. It arranges
it so that the surface density is decreasing and the temperature is increasing with growing WD mass and the width is decreasing, finally.
The width of the thermal BL is increasing with $M_\ast$, on the other hand, although only weakly. This is due to the fact that more
energy is released with increasing WD mass which is distributed on a wider area.
It seems that the dependence on the stellar mass is approximately linear in a logarithmic plot. Thus, as a rule of thumb,
we infer that the width of the BL is decreasing exponentially with increasing WD mass.

\begin{figure}[t]
\begin{center}
\includegraphics[width=.5\textwidth]{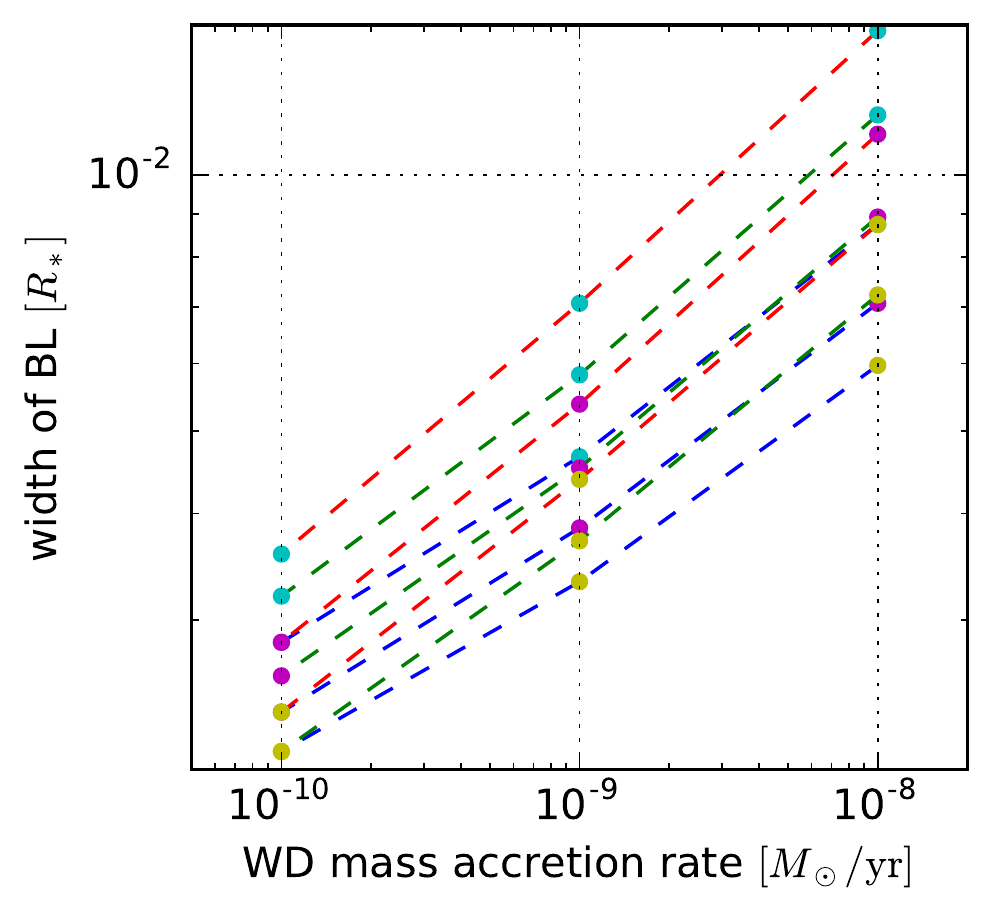}
\caption{\label{fig:width2}%
The width of the dynamical BL as a function of the WD mass accretion rate in units of solar masses per year. The marker color represents
the mass of the WD and cyan, magenta and yellow correspond to $0.6$, $0.8$ and $1.0$ solar masses, respectively.
The line color indicates the rotation rate of the model such that blue,
green and red correspond to $0.2$, $0.7$ and $0.9 \Omega_\mathrm{K}(R_\ast)$, respectively.
}
\end{center}
\end{figure}

Finally, we investigate the dependence on the mass accretion rate of the WD and consider Fig.~\ref{fig:width2} for this purpose. Shown here
are the same nine models as in Fig.~\ref{fig:width}. There is a clear trend for the BL width to grow with increasing mass accretion rate.
Again, this is influenced by the surface density and the temperature which are both growing drastically with increasing mass accretion rate
since a higher $\dot{M}$ means that more mass accumulates in the disk. Accordingly, the energy dissipation through shear is enhanced and
leads to a higher disk temperature. The same holds for the effective temperature and thus the width of thermal BL is growing even more
drastically than the dynamical BL. For high stellar rotation rates (red line), the trend is almost linear in the double logarithmic
plot of Fig.~\ref{fig:width2}. With decreasing stellar rotation rate, however, the trend seems to deviate slightly from linear. Therefore,
while the BL width increases exponentially with the logarithm of the mass accretion rate for high stellar rotation rates, it does only roughly
so for lower rotation rates. We note that the width of the BL is in all cases given in units of the stellar radius, which depends on the
stellar mass.

\subsection{The luminosity of the BL}

\begin{figure}[t]
\begin{center}
\includegraphics[width=\textwidth]{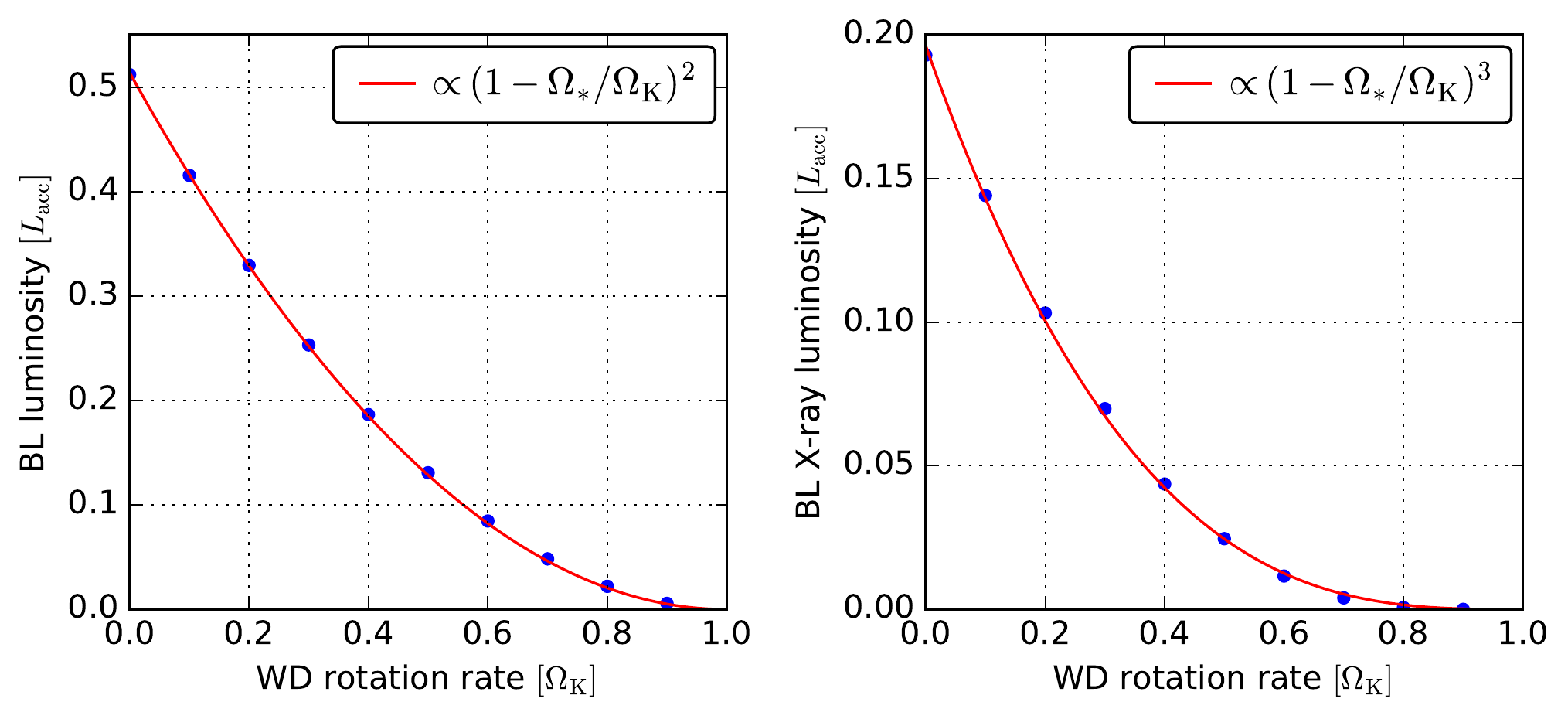}
\caption{\label{fig:lum1}%
The luminosity of the BL as a function of the WD rotation rate. The luminosity is normalized to the total accretion luminosity $L_\mathrm{acc}=GM_\ast\dot{M}/R_\ast$.
The left hand side visualizes the total luminosity of the BL and square fit to the data points. In the right hand side picture, only the X-ray luminosity ($0.1$ to $10$ keV) is
taken into account and plotted along with a cubic fit to the data.
}
\end{center}
\end{figure}

The luminosity of the BL is of great importance when comparing the results of our simulations with real observations. In our model, the BL
and the disk radiate like a black body at each surface point with a temperature that is given by the effective temperature $T_\mathrm{eff}$.
Therefore, the luminosity of one ring is given by
\begin{equation}
	L_i = 2\int_0^{2\pi}\int_{r_i^\mathrm{L}}^{r_i^\mathrm{R}}\sigma T_{\mathrm{eff},i}^4 \ r\,\mathrm{d}r\,\mathrm{d}\varphi,
\end{equation}
where the factor $2$ comes from the two sides of the disk and $r^\mathrm{L,R}$ means the left and the right limit of the individual ring. The
total BL luminosity is then obtained by summing up all rings up to the point where the disk starts. The end of the BL and the beginning of the
disk, respectively, is derived by comparing the effective temperature of our models with the surface temperature given by the standard solution
of the accretion disk by \citet{1973A&A....24..337S}:
\begin{equation}
	T(r) = \left[ \frac{3GM\dot{M}}{8\pi r^3 \sigma} \left( 1 - \left(\frac{R_\ast}{r}\right)^{1/2}\right)\right]^{1/4}\label{eq:SS-temp}
\end{equation}
The disk of our models is perfectly described by the standard solution and especially Eq.~\ref{eq:SS-temp} and thus we define the thermal BL to
extent up to the point where $T_\mathrm{eff}$ comes to within five percent of the standard solution.

Figure \ref{fig:lum1} visualizes the luminosity as a function of the stellar rotation rate for the reference model with $0.8 M_\ast$ and
$\dot{M}=10^{-8} M_\odot/\mathrm{yr}$. With increasing stellar rotation rate, the luminosity of the BL decreases drastically. This is due to the
fact that almost all energy which is released in the BL originates from the difference in kinetic energy of the gas just outside the BL and
at the surface of the WD. The faster the star spins, the less the gas is slowed down and less energy becomes free and contributes to the observed
luminosity. Thirty years ago, there has been a debate about the relation that ties the luminosity of the BL to the stellar rotation rate
\citep[see e.g.][]{1987PhDT.......161K,1991A&A...247...95K,1995ApJ...442..337P}, however, the consent has been to utilize
\begin{equation}
	L_\mathrm{BL} = \frac{1}{2}L_\mathrm{acc}\left(1 - \frac{\Omega_\ast}{\Omega_\mathrm{K}(R_\ast)}\right)^2\label{eq:lum1}
\end{equation}
as a formula. The functional dependence of Eq.~(\ref{eq:lum1}) almost perfectly describes the data as can be seen from the fit (red curve) in
Fig.~\ref{fig:lum1}. However, we found from our simulations that the prefactor of $1/2$ is too low and the fit yielded a value
of $0.51$ for the reference model. This has to do with the definition of the inner disk radius and thus small variations are possible
due to different setups.

\begin{figure}[t]
\begin{center}
\includegraphics[width=\textwidth]{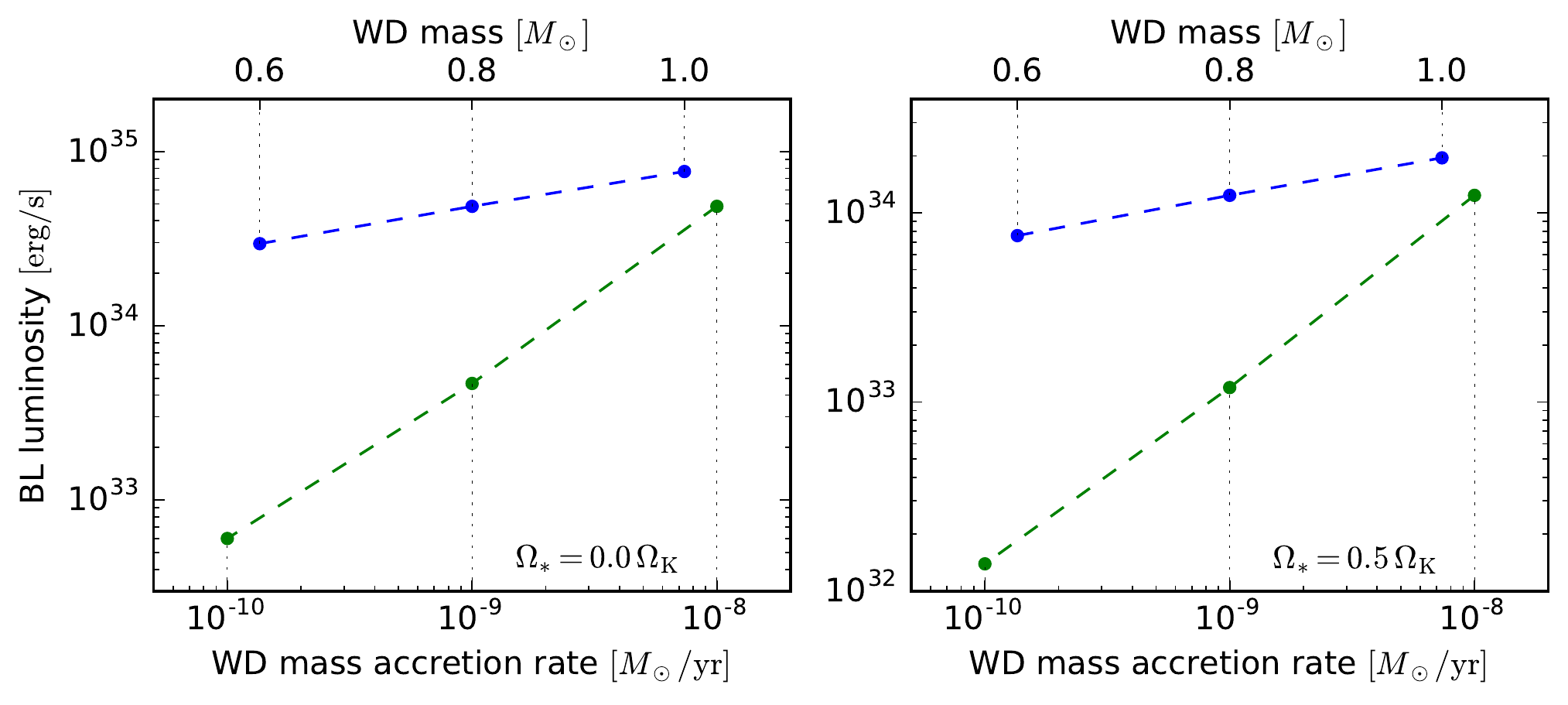}
\caption{\label{fig:lum2}%
The total luminosity of the BL as a function of the stellar mass $M_\ast$ and mass accretion rate $\dot{M}$. The green data points visualize the
luminosity for the three different mass accretion rates $10^{-10}, 10^{-9}$ and $10^{-8} M_\odot/\mathrm{yr}$ and the blue data points correspond
to the mass dependence of the luminosity. The WD is non-rotating
in the left panel, $\Omega_\ast = 0.0\Omega_\mathrm{K}(R_\ast)$, and has a rotation rate of $\Omega_\ast=0.5\Omega_\mathrm{K}(R_\ast)$ in the
right panel.
}
\end{center}
\end{figure}

On the right hand side of Fig.~\ref{fig:lum1} we display the X-ray luminosity of the BL for the energy band between $0.1$ and $10$ keV. In
order to derive $L_\mathrm{BL, X-ray}$, we assume Planck's law for every ring with the temperature $T_\mathrm{eff}(r)$ and integrate over
the solid angle and the disk area to obtain the spectral luminosity of the BL. We can then confine the energy interval of interest and calculate
the luminosity of that band. For our reference model, the X-ray luminosity of the BL around a non-rotating WD amounts to roughly $40\%$
of the total BL luminosity and drops very fast for increasing stellar rotation because only very high effective temperatures
significantly contribute to the X-ray band. The drop is considerably faster than that of the total luminosity and goes with the third
power of the stellar rotation rate. Thus, we postulate as a rule of thumb for the X-ray luminosity of the BL the law:
\begin{equation}
	L_\mathrm{BL, X} = 0.2 L_\mathrm{acc}\left(1-\frac{\Omega_\ast}{\Omega_\mathrm{K}(R_\ast)}\right)^3
\end{equation}
The prefactor might also depend on other parameters of the system. The functional dependence, however, seems to be solid for the cases
we simulated within this study.

In Figure~\ref{fig:lum2}, we verify the dependence of the luminosity on the WD mass and mass accretion rate, which is given by Eq.~(\ref{eq:Lacc}).
The green data points and connecting dashed line refer to the three different mass accretion rates of $10^{-10}, 10^{-9}$ and $10^{-8}$ solar masses
per year. In the double logarithmic plot the data points lie on one straight line which means that indeed $L_\mathrm{BL}\propto \dot{M}$,
independent of $\Omega_\ast$.
Both plots of Fig.~\ref{fig:lum2} differ only in stellar rotation rate. In the left hand side plot, the WD is non-rotating and on the right hand
side it rotates with $50\%$ of the breakup velocity. The data points for the three different WD masses $0.6, 0.8$ and $1.0 M_\odot$ are situated
also on a straight line, though in a semi logarithmic representation. This means that there seems to be a rather weak exponential dependence between
BL luminosity and WD mass. We note that mass and radius are not independent due to the mass-radius relation for WDs. Therefore, the exact dependence
on stellar mass is rather given by the relation $M_\ast/R_\ast$, which depends on the mass-radius relation.
However, for the small mass spectrum observed in WDs, a good first approximation is a linear increase in luminosity with increasing stellar mass.

\begin{figure}[t]
\begin{center}
\includegraphics[width=0.5\textwidth]{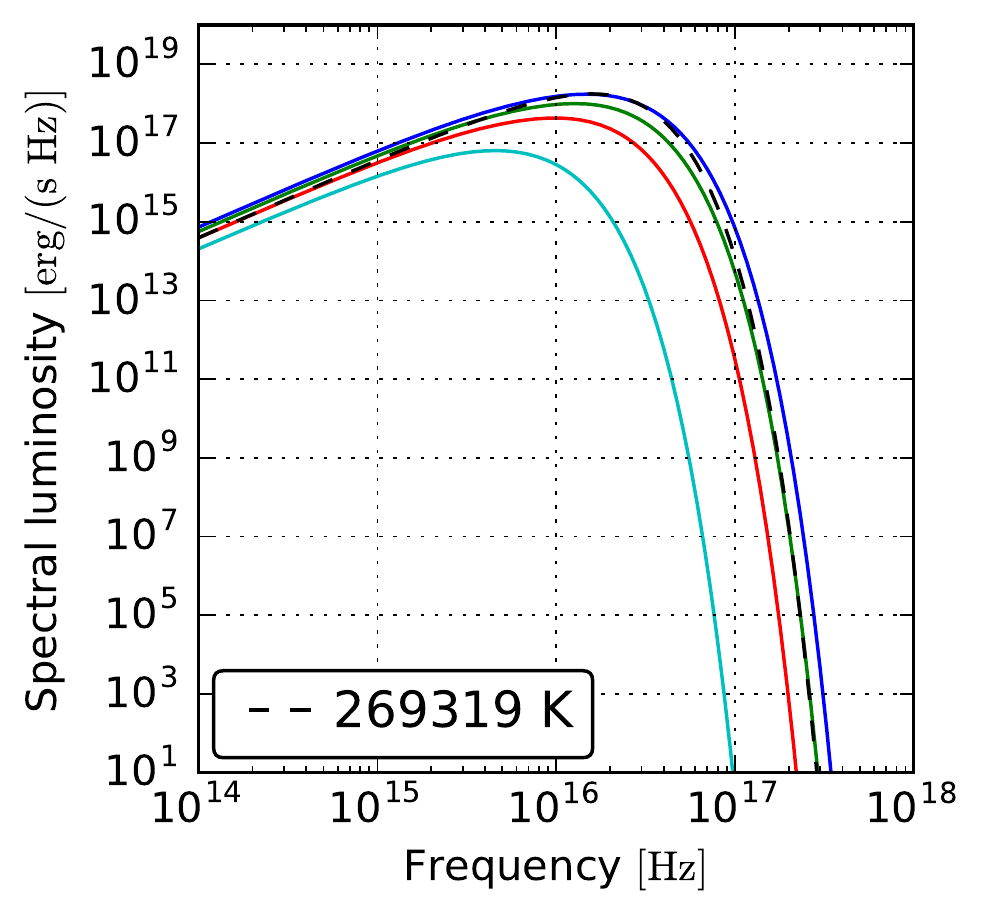}
\caption{\label{fig:spectrum}%
Spectral luminosity $L_\nu$ for the reference model ($0.8M_\odot, 10^{-8}M_\odot/\mathrm{yr}$) and a stellar rotation rate of $0.0$ (blue),
$0.3$ (green), $0.6$ (red) and $0.9\Omega_\mathrm{K}(R_\ast)$ (cyan). The black dashed line indicates a fit with a single temperature Planck
law.
}
\end{center}
\end{figure}

Finally, in Fig.~\ref{fig:spectrum} we show the spectrum of the BL of the reference model which has a $0.8$ solar mass WD and a mass accretion
rate of $10^{-8}$ solar masses per year. The spectral luminosity is obtained by imposing Planck's law with temperature $T_\mathrm{eff}$ for
each ring of the BL and subsequently adding them up while taking into account the area of the ring. In the graph, four different stellar rotation
rates, $0.0$ (blue), $0.3, 0.6$ and $0.9 \Omega_\mathrm{K}(R_\ast)$ (cyan) are presented. The spectrum becomes considerably harder with
decreasing stellar rotation rate which is in accordance with the discussion of the effective temperature shown in Fig.~\ref{fig:Sigma-Teff}.
The high energetic part of the spectrum stems from the hottest parts of the thermal BL, i.e. where the maximum of the effective temperature
is located. The dashed black line in Fig.~\ref{fig:spectrum} indicates a black body fit to the spectrum of the non-rotating WD. To this purpose
we tried to approximate the compound BL spectrum by a single temperature Planck law which gives a rough estimate of the shape of the spectrum.
In this case, the fit yields a black body temperature of approximately $270\,000$ Kelvin. This temperature can then be compared with black body
fits of real observations.

\section{Summary and conclusion}

In this study, we have presented elaborate one-dimensional simulations of the BL around a WD in a cataclysmic variable system. We employed
the thin disk approximation where axisymmetry is assumed and the disk is vertically integrated. The novelty of our model involves the detailed
treatment of radiation in that radiation transport in the radial as well as in the vertical direction has been implemented and that the
radiation energy has been propagated in time along with the thermal energy. For the purposes considered within this study the 1D approximation
is sufficient and, together with observationsi, our results can help to identify system parameters of CVs like SS Aur
(Nabizadeh, Balman, Hertfelder 2017, in preparation). There are, however, questions which involve the structure, the mixing or
instabilities in the BL that can not be answered withing the realm of the 1D model presented here. One must then perform two- or higher
dimensional simulations (e.g. Hertfelder \& Kley 2017, in preparation).

The parameter study we presented here comprises models of the BL around WDs of the masses $0.6, 0.8$ and $1.0$ solar masses and thus covers
the typical mass range of WDs in CVs. For each mass, three mass accretion rates, viz. $10^{-10}, 10^{-9}$ and $10^{-8}$ solar masses per
year, were considered. Finally, for every $M_\ast$-$\dot{M}$ combination, ten stellar rotation rates were imposed, spanning the whole domain
from a non-rotating up to a nearly at break-up velocity rotating WD. Thus, we have run 90 models in total which gave us a good starting
point for a thorough parameter study. As a reference model we took the simulation with $0.8 M_\odot$ and $10^{-8} M_\odot/\mathrm{yr}$
and started by discussing the basic properties of a non-magnetic BL.\footnote{Due to the high number of simulations it is not possible to
include data such as the BL luminosity for each model in this paper. However, numbers not presented here can naturally be requested by
email to the author.}

The BL connects the disk in which the gas is moving with Keplerian velocity with the stellar surface that in general rotates with a lower
velocity and thus the gas looses a good part of its angular momentum and kinetic energy in this region. Our simulations show that the BL
takes the form of a bottleneck in order to accomplish this task: The radial infall velocity increases drastically due to the loss of the
stabilizing angular momentum and simultaneously the surface density drops considerably in order to maintain the constant mass flux. During
the deceleration of the azimuthal velocity component a great deal of energy comes free which is responsible for the sudden rise of the
effective temperature in the BL. Temperatures of up to almost $350\,000$ Kelvin are reached for the non-rotating WD of the reference model.
Since the energy originates from the deceleration process, consequently with increasing stellar rotation the maximum temperature of the
BL decreases more and more rapidly.

The width of the BL is one of the important parameters that we deeper looked into. At first, we have to distinguish between the terminologies
of the dynamical and the thermal BL. The dynamical BL orients itself to the behavior of the angular velocity only and the width is defined
to be the region from the stellar surface up to the maximum of $\Omega(r)$. With increasing WD rotation rate, the width of the dynamical
BL in general increases as well. However, depending on parameters like stellar mass and mass accretion rate, it is not an uncommon phenomenon
that it first shrinks before growing perceptibly. For variations in $M_\ast$ and $\dot{M}$, the situation is clear:
With increasing mass or decreasing mass accretion rate, the dynamical BL becomes considerably thinner. The dynamical width is an
important parameter since it defines the region where the energy which is later radiated away is produced. The more confined this region is,
the harder the radiation will be. However, in this context it is more convenient to refer to the thermal BL. The radiation produced in
the dynamical BL is distributed through radiative diffusion in the radial direction before it can escape from the system. The region where
the heat spreads is denoted as the thermal BL and with a radial extent of $\sim10\%$ of the stellar radius it is roughly ten times as wide
as its dynamical counterpart. In contrast to the dynamical BL, the thermal BL becomes smaller with increasing stellar rotation rate. Its
width directly influences the spectrum of the BL.

The most important parameter when comparing one-dimensional BL simulations with observations is the luminosity. We confirmed that
indeed the BL luminosity decreases quadratically with increasing stellar rotation rate $\Omega_\ast$. 
An interesting result
arises when considering the X-ray band of the BL luminosity only: In this case, we observe a cubic dependence on the stellar rotation
rate. One problem that arises in connection with the BL luminosity is clearly the ambiguity that different models may yield the same total
BL luminosity. For instance, a model with high stellar rotation rate but also a high mass accretion rate can yield the same
luminosity as a model with a smaller mass accretion rate but also a lower rotation rate. Here, one can step in with the additional information
from the X-ray luminosity and utilize the formula we provided for an estimate of the system parameters. Another possibility is given by
regarding the spectrum of the BL which will also help to clear up the ambiguity outlined above.

In the end, we want to learn more about systems that can be observed in space. Thus, by combining the insight gained from theoretical
simulations and observation we should be able to learn more about its parameters. The approach presented here answers
these purposes in that it is elaborated enough to come to unambiguous conclusions by comparing the results with observations, but
at the same time it is still fast enough to run a large number of models which are necessary for the parameter hunt.

%

\bibliographystyle{aa}
\bibliography{references}

\end{document}